\documentclass[12pt,preprint]{aastex}
\usepackage{rotating}
\usepackage{epsfig}
\usepackage{ifpdf}

\begin{document}

\title{Radio monitoring of the periodically variable IR source LRLL 54361: No direct correlation between the radio and IR emissions}

\author{
Jan Forbrich\altaffilmark{1,2}, Luis F. Rodr\'\i guez\altaffilmark{3}, Aina Palau\altaffilmark{3}, Luis A. Zapata\altaffilmark{3}, James Muzerolle\altaffilmark{4} and
Robert A. Gutermuth\altaffilmark{5}}

\altaffiltext{1}{University of Vienna, Department of Astrophysics, T\"urkenschanzstra$\ss$e 17, 1180 Vienna, Austria }

\altaffiltext{2}{Harvard-Smithsonian Center for Astrophysics, 60 Garden St MS 72, Cambridge, MA 02138, USA}

\altaffiltext{3}{Instituto de Radioastronom\'\i a y Astrof\'\i sica, 
UNAM, Apdo. Postal 3-72 (Xangari), 58089 Morelia, Michoac\'an, M\'exico}

\altaffiltext{4}{Space Telescope Science Institute, 3700 San Martin Drive, Baltimore, MD 21218, USA}

\altaffiltext{5}{Department of Astronomy, University of Massachusetts, Amherst, MA 01003, USA}

\email{
 jan.forbrich@univie.ac.at, l.rodriguez,a.palau,l.zapata@crya.unam.mx, muzerol@stsci.edu, rgutermu@astro.umass.edu}
 
\begin{abstract}
LRLL 54361 is an infrared source located in the star forming region IC 348 SW. Remarkably, its infrared
luminosity increases by a factor of 10 during roughly one week every 25.34 days.
To understand the origin of these remarkable periodic variations,
we obtained sensitive 3.3 cm JVLA radio continuum observations of LRLL 54361 and its surroundings in six different
epochs: three of them during the IR-on state and three during the IR-off state.
The radio source associated with LRLL 54361 remained steady and did not show a correlation with the IR variations.
We suggest that the IR is tracing  the results of fast (with a timescale of days) pulsed accretion from an unseen
binary companion, while the radio traces an ionized outflow 
with an extent of $\sim$100 AU that smooths out the variability over a period of
order a year. 
The average flux density measured in these 2014 observations, 27$\pm$5 $\mu$Jy, is about a factor of two less than
that measured about 1.5 years before,  $53\pm$11 $\mu$Jy, suggesting that variability in the radio is present, but
over larger timescales than in the IR. We discuss other sources in the field, in particular two 
infrared/X-ray stars that show rapidly varying gyrosynchrotron emission.

\end{abstract}  


\section{Introduction}

LRLL 54361 is an infrared source (Luhman et al.1998) located in the star forming region IC 348 SW.
It presents a remarkable time variation, with its infrared luminosity increasing 
by a factor of 10 during roughly one week every 25.34$\pm$0.01 days (Muzerolle et al. 2013).
As part of a 3.3 cm study of the IC 348 SW region made with the Karl G. Jansky Very Large Array,
Rodr\'\i guez et al. (2014) detected a source associated with LRLL 54361. 
The infrared variability was attributed to pulsed accretion from an unseen binary companion (Muzerolle et al. 2013).
Since accretion and outflow are expected to be correlated in the current paradigm of star formation (e.g. Pech et al. 2010; Anglada et al. 2015),
we believed that centimeter monitoring of this source could help better understand the origin of
the IR variability.

\section{Observations}
The six observations were made with the Karl G. Jansky Very Large Array of NRAO\footnote{The National 
Radio Astronomy Observatory is a facility of the National Science Foundation operated
under cooperative agreement by Associated Universities, Inc.} centered at the rest frequency of 9.0 
(3.3 cm) GHz during
2014 October 18 and 26, November 10 and 19, and December 3 and 16, under project
14B-251.  At that time the array was in its C configuration.  The phase center was 
at $\alpha(2000) = 03^h~ 43^m~ 51\rlap.^s03$;
$\delta(2000)$ = $+$32$^\circ~ 03'~ 07.7''$, approximately the radio position of LRLL 54361 as determined
by Rodr\'\i guez et al. (2014). In all observations the absolute amplitude calibrator was J1337$+$3309 and
the phase calibrator was J0336$+$3218. 

The digital correlator of the JVLA was configured in 16 spectral windows of 128 MHz width each subdivided 
in 64 channels of spectral resolution of 2 MHz. 
The total bandwidth of the observations was about 2.048 GHz in a full-polarization mode.
The half power width of the primary beam is $\sim 5\rlap{'}.0$ at 3.3 cm. 

The data were analyzed in the standard manner using the VLA Calibration Pipeline\footnote{http://science.nrao.edu/facilities/vla/data-processing/pipeline.} 
 which uses the CASA (Common Astronomy Software Applications) package of NRAO,
although for some stages of the analysis we used the AIPS (Astronomical Image Processing System)
package. 
For all the imaging, we used the ROBUST parameter of CLEAN set to 2 (Briggs 1995), to obtain a better sensitivity, at the expense of losing some 
angular resolution. All images were corrected for the response of the primary beam, increasing the noise away from the phase center.
The resulting rms of the averaged image (over all six epochs) was  3 $\mu$Jy beam$^{-1}$ at the center of the field, with an angular resolution 
of $3\rlap.{''}1 \times 2\rlap.{''}7$ with PA = $-69^\circ$. In Table 1 we give the positions, flux densities and deconvolved sizes of the sources
detected, as obtained from the
average image. With the exception of sources JVLA 3a, JVLA 3b, and JVLA 4 that show marginal evidence of being slightly extended
(see Table 1), all other sources appear unresolved. The rms of the individual images at the center of the field was $\sim$7 $\mu$Jy beam$^{-1}$ . 

\section{Comments on individual sources}

\subsection{JVLA 5 and JVLA 6: two rapidly variable gyrosynchrotron sources}

The only two sources that are clearly detected in the individual images at some epochs are JVLA 5 (infrared source IC 348 LRL 49)
and JVLA 6 (infrared source IC 348 LRL 13), both of which have been detected before. We follow here the nomenclature of Rodr\'\i guez et al. (2014) for the VLA sources.
In Table 2 we show the flux densities of these two sources at each epoch of observation, that are seen to vary over an order of magnitude in timescales
of days. In Figure 1 we show the contour image of these two sources, from the data averaged over the six epochs.
These two stars are the only objects reported here that coincide with Chandra X-ray sources, CXOUJ034357.62+320137.4  
(JVLA5) and CXOUJ034359.67+320154.1 (JVLA6), as reported in the study of
Stelzer et al. (2012). They are also two out of four sources in this region that were detected simultaneously in X-ray and radio emission by Forbrich et al. (2011). Our results support the interpretation of Rodr\'\i guez et al. (2014)
that these two radio sources are associated with young stars with active magnetospheres,
such as those detected in other regions of star formation (e.g. Dzib et al. 2013; Liu et al. 2014).
This class of extremely compact radio sources has been very useful for the accurate
determination of the parallax (and distance) to several regions of star formation (e.g. Loinard et al. 2011).

\subsection{JVLA 1 (=LRLL 54361) and JVLA 10 }

JVLA 1 is the radio counterpart to the variable IR source LRLL 54361 and the main target
of the observations reported here. It is associated with a jet-like feature to its NW that is seen in narrowband $H_2$ (2.12 $\mu$m;
Walawender et al. 2006) and 1.6 $\mu$m continuum (Muzerolle et al. 2013) imaging. An elongated structure is also seen to
the NE in these infrared images.
In a combined X-ray and cm radio study of the region, Forbrich et al. (2011) did not find evidence for either an X-ray or a radio source toward LRLL~54361 using \textit{Chandra} and the "classic"
NRAO VLA on March 13 and 18, 2008. The radio sensitivity was 20~$\mu$Jy (5$\sigma$) per epoch in both X- and C-band. Coincidentally, the first of these observations fall onto a peak of an infrared pulse in an epoch that was observed with \textit{Spitzer}. However, in JVLA observations in the C configuration carried out in June 2013, Rodr\'iguez et al. (2014) detected the source with a total flux density of $53.3\pm3.1$~$\mu$Jy (at 9.0~GHz) and a flat spectral index of $\alpha = -0.3\pm 0.2$ (with observations at 9~GHz and 14 GHz). At the same flux level, the source thus should have been clearly detected in 2008 at S/N$\sim$13, suggesting variability. The 2013 observations were carried out about 2 days after a pulse peak. The relation of the radio flux density and the phase of the infrared pulse variability was therefore inconclusive, and we carried out targeted on-peak and off-peak observations.

In our average image including all six new epochs (see Figure 2), LRLL 54361
has a 3.3 cm flux density of 
27$\pm$4 $\mu$Jy, about one half of the value given from the 2013 observations of Rodr\'\i guez et al. (2014), 53$\pm$11 $\mu$Jy. 
We searched for evidence of direct correlation with the IR variability by making one image with the three IR-on epochs
(Oct. 18, Nov. 14 and Dec. 03 of 2015) and another with the three IR-off epochs (Oct. 26, Nov. 10 and Dec. 16 of 2015). 
We obtain similar flux densities in these two averaged images,
23$\pm$5 $\mu$Jy for the IR-on epochs and 27$\pm$5 $\mu$Jy for the IR-off epochs. 
The VLA observations of
Forbrich et al. (2011) were taken on 2008 March 13 and 18, with the latter epoch coinciding with an IR peak.
The upper limits at the position of LRLL 54361 in the concatenated data at X- and C-bands were in the order
of 0.08 mJy (5-$\sigma$).
We conclude that there is no direct correlation between the
radio and IR emissions of this source. Finally, this source has no reported X-ray counterpart
in the observations of  Stelzer et al. (2012).

One possible explanation for this lack of correlation between the IR and radio emissions could be attributed to the timescale of the
variations. The IR emission is probably tracing phenomena that take place in the accretion disk close to the central star. Given the IR variability 
observed, the timescale of these phenomena must be a few days or less. On the other hand, a power-law fit to the model of
Muzerolle et al. (2013) for the expected dust flux density near maximum emission
at millimeter and centimeter wavelengths gives:

$$\Biggl[{{S_\nu} \over {\mu Jy}}\Biggr] \simeq 3.5 \Biggl[{{\nu} \over {10~GHz}} \Biggr]^{2.9}.$$

We then expect a flux density from the dust of $\sim$3 $\mu$Jy at 9 GHz. Since we observe a flux density an order of magnitude larger ($\sim$27 $\mu$Jy),
we propose that the ``excess'' centimeter continuum emission is most
probably tracing an ionized jet, as observed in several other sources (e.g. Rodr\'\i guez et al. 1998). 
These jets have typical detectable dimensions in the order of $\sim$100 AU (Anglada et al. 2015). Assuming a jet velocity of order
200 km s$^{-1}$, this outflow is averaging the ejections over a timescale in the order of one year and smoothing out possible faster
variations directly associated with the IR variability. The lack of fast, significant radio variability in JVLA 1 also argues against an
explanation in terms of gyrosynchrotron emission from active magnetospheres (e.g. Torres et al. 2012)  or synchrotron emission from colliding magnetospheres
(Salter et al. 2010) for the centimeter source.

A lack of correlation between the infrared and radio variable emissions has also been observed for other types of objects.
Recently, the infrared source HOPS 383 was reported to have a mid-infrared -- and bolometric -- luminosity increase 
of a factor of 35 between 2004 and 2008
(Safron et al. 2015), becoming the first clear example of a class 0 protostar  with a large accretion burst. 
However, Galv\'an-Madrid et al. (2015) showed that in the time period  between 1998 and 2014
the 3.0 cm flux density of the source varied only mildly, staying at a level
between 200 and 300 $\mu$Jy. These authors interpret the absence of a radio burst as implying that accretion and ejection
enhancements do not follow each other in time, at least not within timescales shorter than a few years.

JVLA 10 is a new detection (see Figure 2). It coincides within 1" with
LRLL 54362 and appears located about 16" to the north of JVLA 1 (= LRLL 54361). 
JVLA 10 is also associated with the young stellar object IC 348 3 (Evans et al. 2009; Gutermuth et al. 2009). In the \sl Spitzer \rm 
IRAC 4.5 $\mu$m image shown in Figure 1 of Rodr\'\i guez et al. (2014) JVLA 10 appears to be associated with the head of an infrared
cometary nebula to its north. This cone-shaped nebula is clearly seen in the HST images analyzed by Muzerolle et al. (2013), but is not shown in their paper.

\subsection{The core of the HH 797 outflow}

In this zone we detect the same sources reported by Rodr\'\i guez et al. (2014). JVLA 3a and JVLA 3b are two sources separated by
$\sim$3$''$ (see Figure 3) and located at the center of the HH 797 bipolar outflow (Pech et al. 2012). 
While in the
observations of Rodr\'\i guez et al. (2014) the source 
JVLA 3b was brighter than the source JVLA 3a, the roles have reversed in these new
observations, indicating variability on timescales of one year. JVLA 3a has LRLL 57025 as its K (2.2 $\mu$m) band counterpart.

JVLA 3c is an interesting
source associated with the submillimeter source 
IC348-SMM2E and with a Class 0 proto-brown dwarf candidate (Palau et al. 2014). Combining the 14.0 GHz
observations of Rodr\'\i guez et al. (2014) with those at 9.0 GHz reported here, we derive a spectral index of
0.4$\pm$0.8 for IC348-SMM2E, consistent with a thermal jet nature, but unfortunately the large uncertainty
prevents from a more definitive characterization. Faint centimeter continuum emission has
been reported for a handful of other proto-brown dwarf candidates (Morata et al. 2015).

The nature of JVLA 4 remains undetermined. Except for a two-band infrared detection reported as part of the c2d Spitzer final 
data release (Evans et al. 2003), it has no known counterpart at other wavelengths. The flux density reported here 
is 1.5 times larger than that measured in 2013. However, the associated errors are large and the source could as well
be steady.

\subsection{New sources}

As discussed above, JVLA 10 is a newly detected radio source that is associated with the young stellar object IC348 3.
Another new source is JVLA 9, at the western edge of the field of view. This source has an infrared counterpart that was flagged as a candidate galaxy by Evans et al. (2003). Its detection in this experiment was favoured by the fact that the
phase center of these observations was $\sim$76$''$ to the west of those of Rodr\'\i guez et al. (2014). 
By the same reason, we did not detect here source JVLA 8, probably because it is far to the east from the new phase center.

\begin{figure}
\centering
\includegraphics[angle=0,scale=0.6]{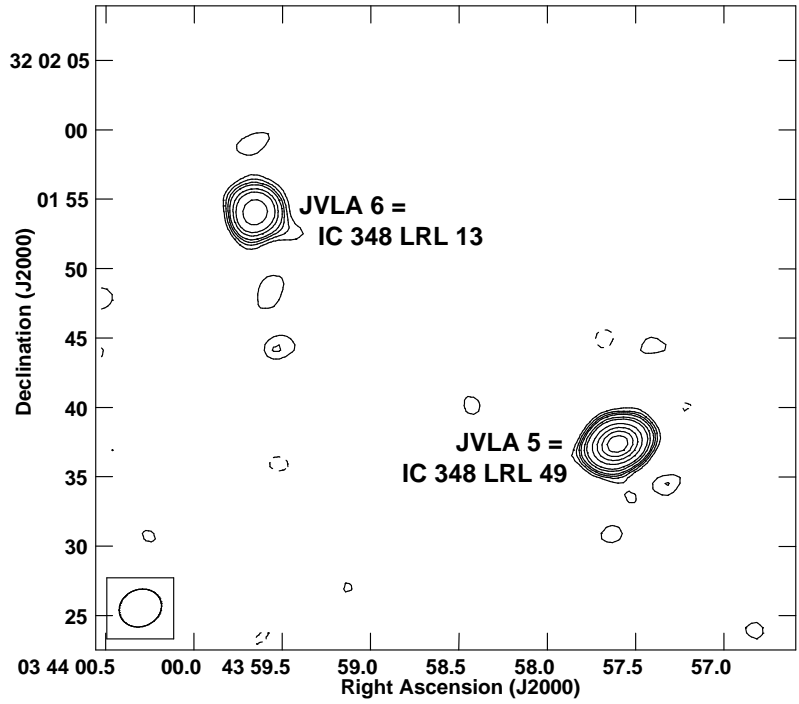}
\caption{\small JVLA 3.3 cm  continuum contour image of JVLA 5 and JVLA 6.
The contours are -4, -3, 3, 4, 5, 6, 8, 10, 15, 20, 25, 30, and 35 times 5.4
$\mu$Jy beam$^{-1}$, the rms noise of this region of the
image. The image has been corrected for the response of the primary beam
and the noise increases away from the phase center. 
The half-power contour of the synthesized beam is shown in the bottom left corner.}
\label{fig1}
\end{figure}

\pagebreak

\begin{figure}
\centering
\includegraphics[angle=0,scale=0.6]{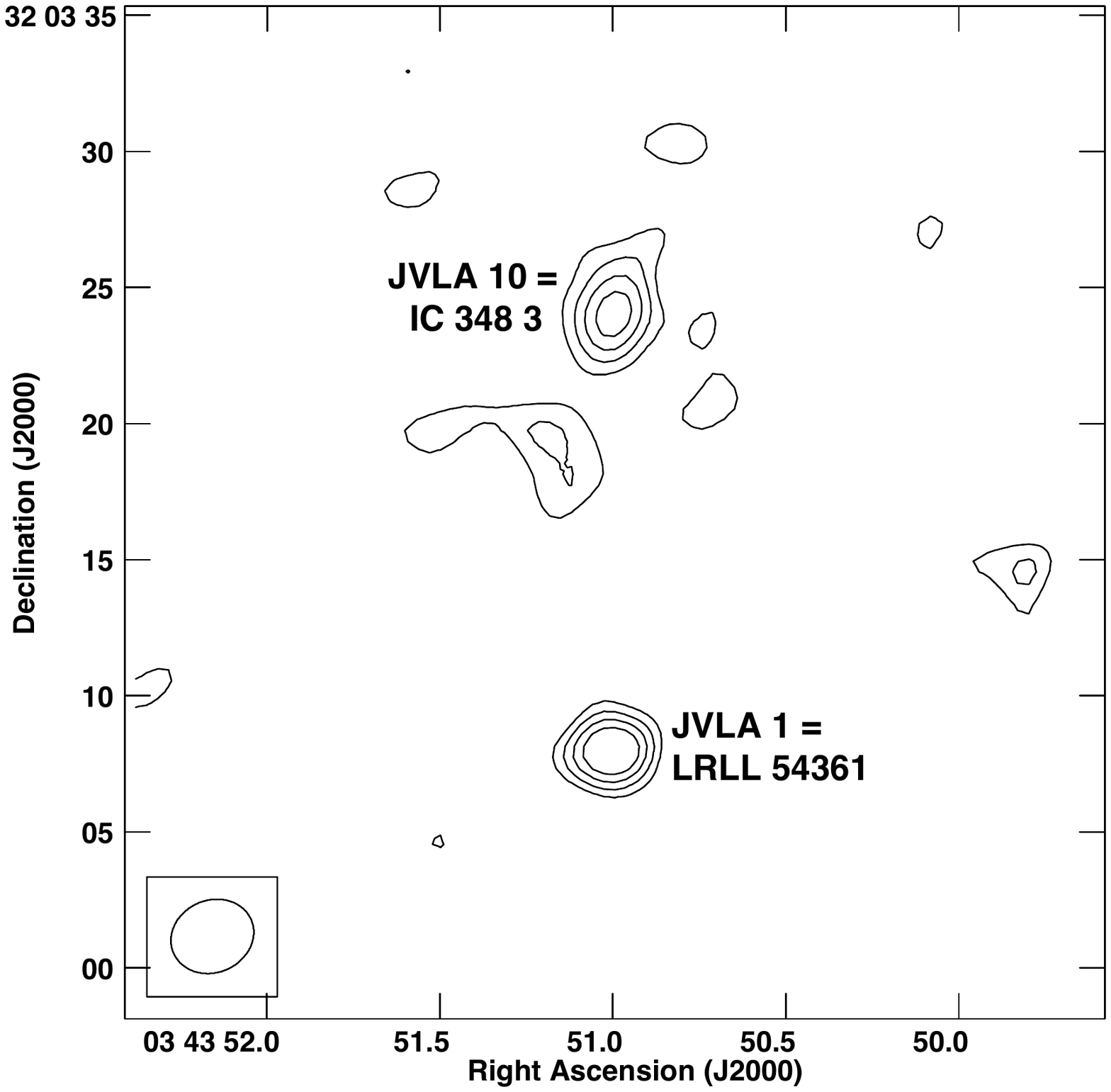}
\caption{\small JVLA 3.3 cm  continuum contour image of JVLA 1 and JVLA 10.
The contours are -4, -3, 3, 4, 5, and 6 times 3.6
$\mu$Jy beam$^{-1}$, the rms noise of this region of the
image. The image has been corrected for the response of the primary beam
and the noise increases away from the phase center. 
The half-power contour of the synthesized beam is shown in the bottom left corner.}
\label{fig2}
\end{figure}

\pagebreak

\begin{deluxetable}{l c c c c c c c c}
\tabletypesize{\scriptsize}
\tablecaption{Parameters of the JVLA sources detected at 3.3 cm in the averaged image}
\tablehead{                        
\colhead{}                        &
\colhead{}                        &
\multicolumn{2}{c}{Position} &
\colhead{}                              &
\multicolumn{3}{c}{Deconvolved size$^b$} &        \\
\colhead{}   &
\colhead{}   &
\colhead{$\alpha_{2000}$}          &
\colhead{$\delta_{2000}$}           &
\colhead{Flux Density$^a$ }       &                            
\colhead{Maj.}  &
\colhead{Min.}  &
\colhead{P.A.}  & \\
\colhead{Source}                              &
\colhead{Name}                              &
\colhead{(h m s) }                     &
\colhead{($^\circ$ $^{\prime}$  $^{\prime\prime}$)}              &
\colhead{($\mu$Jy)}  & 
\colhead{($^{\prime\prime}$)}  &
\colhead{($^{\prime\prime}$)}  &
\colhead{($^\circ$)} &
}
\startdata

JVLA 9     &    -                 & 03 43 42.094  & $+$32 02 25.29 & 35 $\pm$  6 &  - &  - & -\\
JVLA 10   &   IC348 3  &  03 43 50.998  & $+$32 03 24.16 & 18 $\pm$ 3  &  - &  - & -\\
JVLA 1     &   LRLL 54361  & 03 43 51.007 & $+$32 03 08.02 &  27 $\pm$ 5 &  - &  - & -\\
JVLA 2     &   HH 211-MM   & 03 43 56.790 & $+$32 00 50.19   &  92 $\pm$ 6 &  - &  - & - \\
JVLA 3a  &  HH 797-SMM2  & 03 43 56.887 & $+$32 03 03.10  & 62 $\pm$ 14 &  2.7 $\pm$ 1.7 &  $\leq$3.9 & 88 $\pm$ 39 \\
JVLA 3b   &  HH 797-SMM2  & 03 43 57.064 & $+$32 03 05.08  &  37  $\pm$ 12 &  $\leq$2.8&  $\leq$2.7 & 89 $\pm$ 26\\
JVLA 3c   &   IC348-SMM2E      & 03 43 57.740 & $+$32 03 10.37   &  23  $\pm$ 4 &  -&  - & -\\
JVLA 4     &    -   & 03 43 57.088 & $+$32 03 29.78   &   72 $\pm$ 19 &  4.2 $\pm$ 1.7 &  $\leq$3.0 & 130 $\pm$ 31\\
JVLA 5     &   IC 348 LRL 49          & 03 43 57.604 & $+$32 01 37.37   &  187 $\pm$ 6 &  -&  - & - \\
JVLA 6     &   IC 348 LRL 13          & 03 43 59.651 & $+$32 01 54.07 &  101 $\pm$ 6 &  -&  - & -\\
JVLA 7     &    -                               & 03 44 01.640 & $+$32 04 39.72   &   58 $\pm$ 10 &  -&  - & -\\
\enddata
\tablecomments{
                (a): Total flux density corrected for primary beam response. Upper limits are at the 4-$\sigma$ level.\\
                (b): These values were obtained from the task JMFIT of AIPS.}
\end{deluxetable}

\pagebreak

\begin{deluxetable}{l c c c}
\tabletypesize{\scriptsize}
\tablecaption{Flux densities$^a$ at 3.3 cm of the two time variable sources}
\tablehead{                        
\colhead{   }                              &
\colhead{LRL 49}                              &
\colhead{LRL 13}                     &   \\
\colhead{Epoch}                              &
\colhead{($\mu$Jy)}                              &
\colhead{($\mu$Jy)}                     &
}
\startdata

2014 Oct 18   &  116$\pm$10    & 44$\pm$11  \\
2014 Oct  26 &  234$\pm$11   &   $\leq$48 \\
2014 Nov  10  &   971$\pm$11  & 112$\pm$13 \\
2014 Nov 19    &   118$\pm$12   & 266$\pm$13  \\
2014 Dec 03   &  142$\pm$12  &  $\leq$39 \\
2014 Dec 16   &  39$\pm$9  & 56$\pm$13 \\

\enddata
\tablecomments{
                (a): Total flux density corrected for primary beam response. Upper limits are at the 4-$\sigma$ level.}
\end{deluxetable}

\pagebreak

\begin{figure}
\centering
\includegraphics[angle=0,scale=0.6]{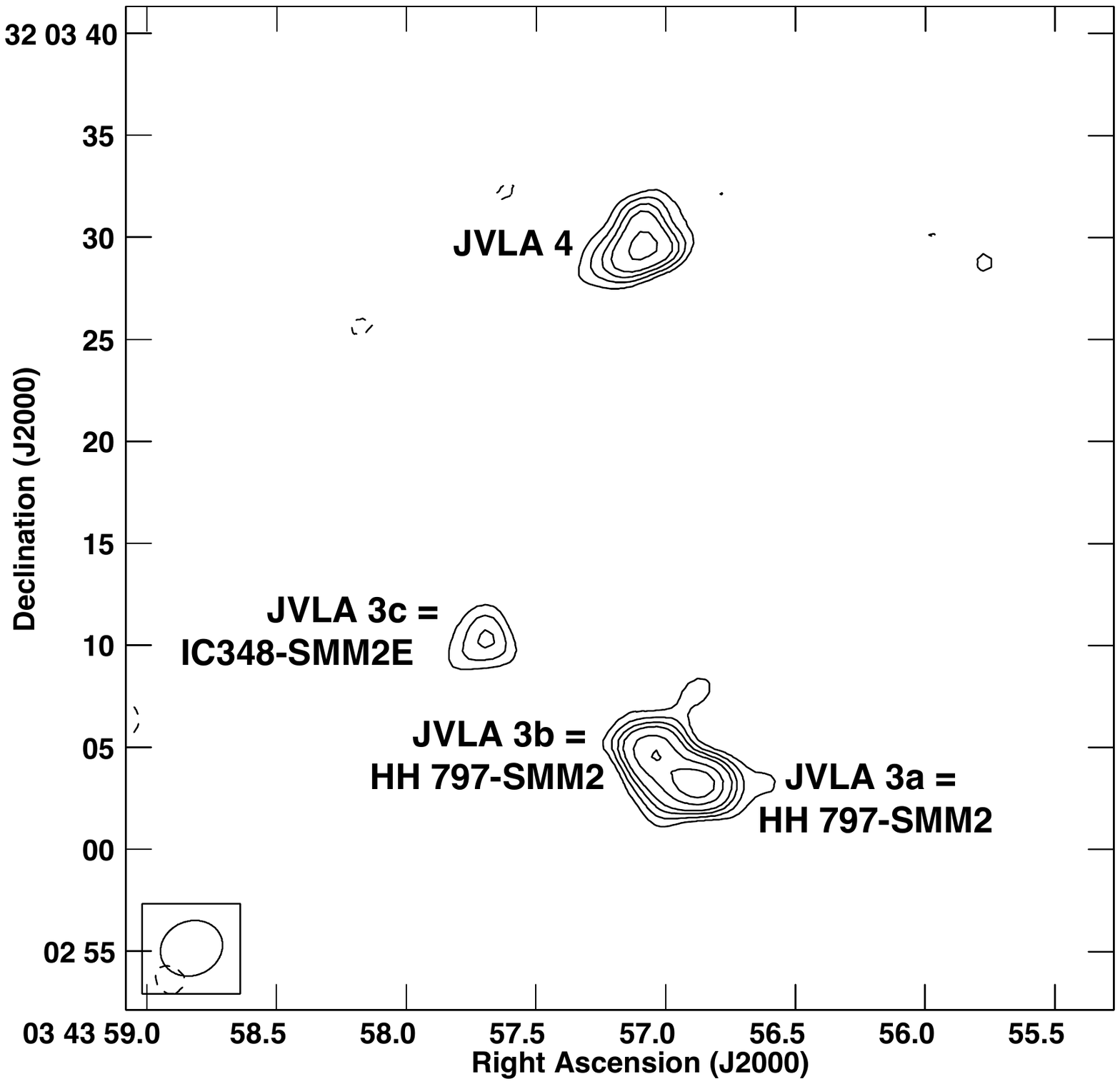}
\caption{\small JVLA 3.3 cm  continuum contour image of JVLA 3 and JVLA 4, at the core of
the HH 797 outflow.
The contours are -4, -3, 3, 4, 5, 6, 8 and 10 times 4.2
$\mu$Jy beam$^{-1}$, the rms noise of this region of the
image. The image has been corrected for the response of the primary beam
and the noise increases away from the phase center. 
The half-power contour of the synthesized beam is shown in the bottom left corner.}
\label{fig3}
\end{figure}

\pagebreak

\section{Conclusions}

The high sensitivity of the Jansky VLA allows the detection of new, previously undetectable faint sources in
regions of star formation. The main results of this new 3.3 cm study of IC 348 SW 
can be summarized as follows.

1. We detected a total of 11 compact radio sources, determining their positions and
flux densities. Only two are new detections.

2. The source JVLA1 is associated with the remarkable periodic time-variable infrared source
LRLL 54361 (Muzerolle et al. 2013). Our monitoring of the region in six epochs shows no
correlation between the radio and IR emissions. We suggest that the radio emission is probably tracing
outflow phenomena that are averaged over timescales of years and cannot detect the rapid variability
found in the IR.

3. A determination of the radio spectral index of the proto-brown dwarf IC348 SMM2E is consistent with a thermal
jet nature.

4. As proposed previously, two of the sources (JVLA5 and 6) are each associated with infrared/X-ray young stars
and most probably are gyrosynchrotron emitters, useful for future high-accuracy astrometric work.

\acknowledgments

Part of this work was made while JF stayed at UNAM as a Visiting Professor under the PAEP 
program. This research has made use of the SIMBAD database,
operated at CDS, Strasbourg, France.
LFR and LAZ are grateful to CONACyT, Mexico and DGAPA, UNAM for their financial
support. AP acknowledges financial support from UNAM-DGAPA-PAPIIT IA102815 grant, M\'exico.

\clearpage

\end{document}